\documentstyle[twoside,fleqn,espcrc2,epsf]{article}

\title{Finite-Temperature Phase Structure of Lattice QCD with the Wilson Quark
Action for Two and Four Flavors}

\author{S. Aoki\address{Institute of Physics, University of Tsukuba,
Tsukuba, Ibaraki 305, Japan}, T. Kaneda$^{\rm a}$, A. Ukawa$^{\rm a}$, T.
Umemura$^{\rm a}$}

\begin{document}

\begin{abstract}
We present further analyses of the finite-temperature phase structure of
lattice
QCD with the Wilson quark action based on spontaneous breakdown of
parity-flavor symmetry.  Results
are reported on (i) an explicit demonstration of spontaneous breakdown of
parity-flavor symmetry beyond the critical line, (ii) phase structure and
order of chiral transition for the case of $N_f=4$ flavors, and (iii) approach
toward the continuum limit.
\end{abstract}

\maketitle
\section{Introduction}

Elucidating the finite-temperature phase structure of lattice QCD with the
Wilson
quark action is complicated by the fact that the action explicitly breaks
chiral
symmetry\cite{earlywork}.  Last year we reported an analysis of this
problem\cite{auu} based on the idea of spontaneous breakdown of
parity and flavor symmetry\cite{aoki}.  The main findings, obtained through
simulations of the two-flavor system on an $8^3\times 4$ lattice, were that (i)
the critical line, defined as the line of vanishing pion screening mass for
finite temporal lattice sizes, does not extend to arbitrarily weak coupling,
but
turns back toward strong coupling at a finite value of $\beta$, forming a cusp
on
the $(\beta, K)$ plane, and (ii) the thermal line of finite-temperature
transition $K=K_t(\beta)$  runs past the tip of the cusp without crossing the
critical line.

We have also argued, based on analytical considerations including those of the
two-dimensional Gross-Neveu model, that the cusp will move toward weak coupling
as the temporal size $N_t$ increases, pushing the thermal line in front
and eventually pinching it at $K=1/8$ and $\beta=\infty$ as $N_t\to\infty$.  In
our view it is only in this limit that the true chiral phase transition
emerges.

While this view provides a consistent picture on how
chiral phase transition arises for the Wilson quark action, we have left
unanswered several important questions to be clarified.  These
are (i) an explicit demonstration that parity-flavor symmetry is spontaneously
broken inside the cusp, (ii) how the phase diagram depends on the number of
flavors $N_f$, in particular if the chiral transition is of first order for
$N_f\geq 3$ as predicted by the sigma model analysis, and (iii) how the tip of
the cusp moves for an increasing temporal lattice size.  In this article, we
present results of our study on these questions carried out since last year.

\section{Evidence for spontaneous breakdown of parity-flavor symmetry}

Spontaneous breakdown of parity-flavor symmetry inside the cusp is signaled by
a non-vanishing vacuum expectation value of the pion field.  For the case of
two
flavors $N_f=2$ which we analyze,  one may take the pion condensate to
point in the
$\tau_3$ direction
$<\overline{\psi}i\gamma_5\tau_3\psi>\ne 0$.   The
pion spectrum will then consist of a massive $\pi^0$ and massless $\pi^\pm$
which are the Nambu-Goldstone modes of the broken symmetry.

In order to ascertain these predictions we carry out hybrid Monte Carlo (HMC)
simulations,  adding a symmetry-breaking term
$\delta S_W=2KH\sum_n\overline{\psi}_ni\gamma_5\tau_3\psi_n$ to the action
to  avoid infrared divergences due to massless $\pi^\pm$
modes.

Runs are made on an $8^3\times 4$ lattice at $\beta=3.5$ with
$0.21\leq K\leq 0.28$ in steps of $0.01$.  The range for the hopping parameter
includes the interval
$K_c=0.2267(2)\leq K\leq 0.2463(7)=K_c^\prime$ previously estimated to be the
parity broken phase\cite{auu}.  For the external field we take $H=0.2$, 0.1,
0.05, 0.02 and 0.01. For each value of $K$ and $H$, typically
$20-50$ trajectories with 0.5 time units are made for thermalization  followed
by 50 trajectories for measurements.  The presence of external field
significantly reduces fluctuations, and we find these statistics to be
sufficient. Hadron screening masses are calculated by periodically
doubling the lattice in the  spatial directions.

\begin{figure}[t]
\centerline{\epsfysize=45mm \epsfbox{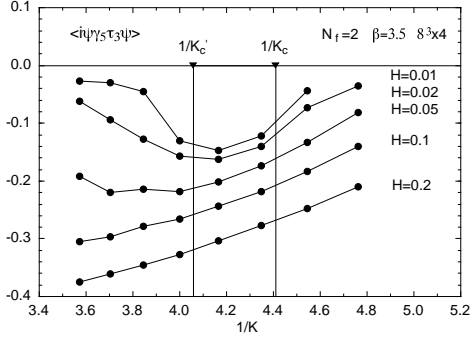}}
\vspace*{-10mm}
\caption{Parity-flavor order parameter as a function of $1/K$ for various
values of external field $H$ at $\beta=3.5$ on an $8^3\times 4$ lattice
for $N_f=2$.}
\vspace*{-8mm}
\label{fig:fig1}
\end{figure}

In fig.~\ref{fig:fig1} we plot the parity-flavor order parameter
$<\overline{\psi}i\gamma_5\tau_3\psi>$ as a function of $1/K$ for $H=0.2-0.01$.
Vertical lines mark the position of the critical lines at $\beta=3.5$.  We
clearly observe that the order parameter inside the cusp ({\it
i.e.}, between the two critical values $K_c$ and $K_c^\prime$) tends to a
non-vanishing value as the external field
$H$ is reduced, while it decreases toward zero outside.

\begin{figure}[t]
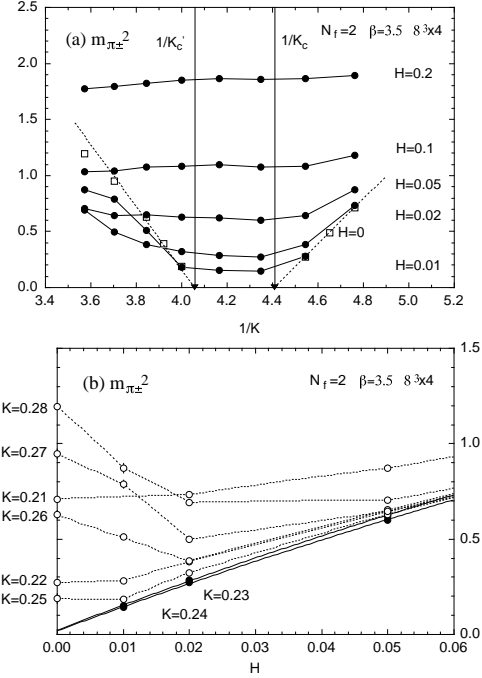

\centerline{\epsfysize=45mm \epsfbox{fig2a.epsf}}
\centerline{\epsfysize=45mm \epsfbox{fig2b.epsf}}
\vspace*{-10mm}
\caption{(a) $m_{\pi^\pm}^2$ as a function of $1/K$ for
various values of external field $H$ at $\beta=3.5$ on an $8^3\times 4$ lattice
for the case of $N_f=2$. Open squares are results for $H=0$. (b) Same as (a)
plotted as a function of $H$ for $H\leq 0.06$.}
\vspace*{-8mm}
\label{fig:fig2}
\end{figure}

The $\pi^\pm$ screening mass squared is plotted in
fig.~\ref{fig:fig2}(a) as a function of $1/K$ and in fig.~\ref{fig:fig2}(b) as
a function of $H$. Inside the cusp
$m_{\pi^\pm}^2$ decreases toward zero, while outside it converges toward values
calculated with $H=0$(open squares).  For $K=0.23$ and 0.24, which are inside
the cusp, lines in fig.~\ref{fig:fig2}(b) show quadratic fits of form
$m_{\pi^\pm}^2=A+BH+CH^2$.  A small non-vanishing value of the intercept
$A\approx 0.02$ is ascribed to finite spatial size effects.

We conclude that the behavior of both the parity-flavor order
parameter and
$m_{\pi^\pm}^2$ strongly supports spontaneous breakdown of parity-flavor
symmetry inside the cusp.

An interesting
point to note is that $m_{\pi^\pm}^2$ converges to the limit $H=0$ from below
for $K>K_c^\prime$, while the approach is from above for $K<K_c$.  In other
words, the parity-flavor broken phase is enlarged beyond the upper part of the
critical line in the presence of the external field.

We also mention that extracting the $\pi^0$ mass inside the cusp requires
computation of disconnected quark loop contribution in the $\pi^0$
propagator.  This may be carried out with the technique of wall source without
gauge fixing previously applied to the $\eta^\prime$
propagator\cite{kuramashi}.
We leave this interesting problem for future work.

\section{Phase diagram for the case of $N_f=4$}

Our analysis of the $N_f=4$ system is made for the dual purpose of
confirming the cusp structure of the critical line and examining the dependence
of the order of chiral transition on
$N_f$.

HMC runs with $H=0$ are carried out on an $8^3\times 4$
lattice at
$0\leq\beta\leq 4.0$ and $0.19\leq K\leq 0.30$.  Thermalization of $20-50$
trajectories with 0.5 time units followed by
$50-100$ trajectories for measurements are made at each $\beta$ and $K$,
chaining runs in $K$ for each value of
$\beta$.  The location of the critical line is estimated from results of
$m_\pi^2$ calculated on a spatially doubled lattice,
and the position and order of  the thermal
transition is examined through behavior of physical quantities
and their time histories.

\begin{figure}
\centerline{\epsfxsize=70mm \epsfbox{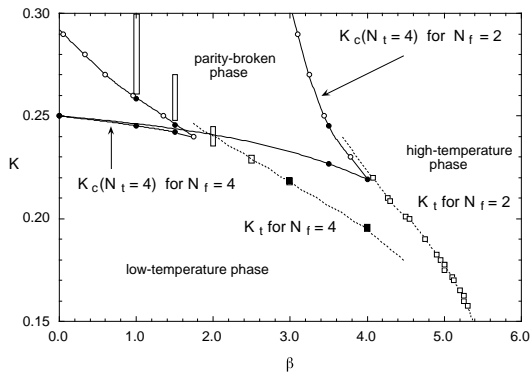}}
\vspace*{-10mm}
\caption{Phase diagram for the $N_f=4$ system for $N_t=4$.  Previous
result for $N_f=2$ is also shown for comparison.  Solid lines are
critical lines and dotted lines the line of thermal transition.  Solid
squares mark points where first-order signals are found, while
open rectangles represent region of smooth crossover.}
\label{fig:fig3}
\vspace*{-8mm}
\end{figure}

In fig.~\ref{fig:fig3} we show the phase diagram for $N_f=4$
together with that for $N_f=2$ previously reported\cite{auu}.  We find a cusp
structure similar to that of $N_f=2$ except for a shift of
$\delta\beta\approx 1.2$ toward stronger coupling, which  we qualitatively
expect from a larger magnitude of sea quark effects for
$N_f=4$.  We also find strong first-order signals across the thermal line
away from the tip of the cusp as marked by solid squares.  Surprisingly,
however,  the transition becomes weaker toward the tip of the cusp, apparently
turning into a smooth crossover at
$\beta=2.5-2.0$.  The location of the crossover is indicated by open rectangles
in  fig.~\ref{fig:fig3}.

\begin{figure}[t]
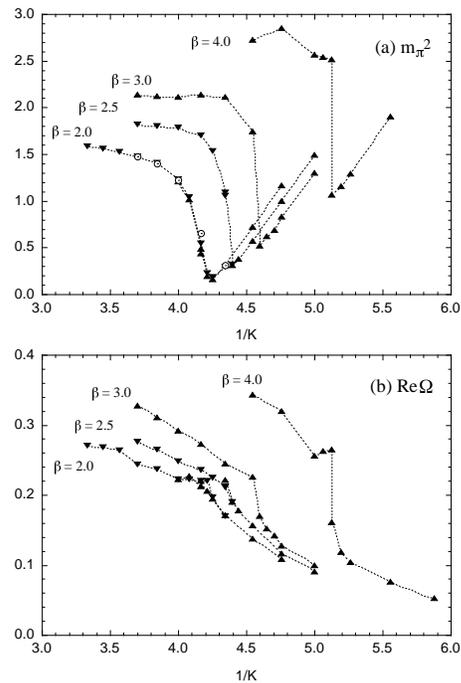

\centerline{\epsfysize=45mm \epsfbox{fig4a.epsf}}
\centerline{\epsfysize=45mm \epsfbox{fig4b.epsf}}
\vspace*{-10mm}
\caption{(a) $m_\pi^2$ as a function of $1/K$ for
various values $\beta$ on an $8^3\times 4$ lattice
forthe case of $N_f=4$. Open circles for $\beta=2.0$ are taken on an
$12^3\times 4$ lattice.  (b) Same as (a) for the real part of Polyakov loop
$Re\Omega$.}
\vspace*{-8mm}
\label{fig:fig4}
\end{figure}

The smoothing of the transition close to the cusp is illustrated
in fig.~\ref{fig:fig4}(a) for $m_\pi^2$ and in (b) for the Polyakov line
$\Omega$.  Jumps in physical quantities are apparent at $\beta=4.0$ and 3.5.
However, at
$\beta=2.5$ and 2.0, data taken for increasing and  decreasing values of
$K$ overlap with each other and do not show any sign of discontinuity.

At $\beta=2.0$ runs are also made with an increased spatial size of $12^3\times
4$.  Results for $m_\pi^2$ plotted by open circles do not show deviation from
those on an $8^3\times 4$ lattice.  Thus it is unlikely that finite-size
effects
has rounded a first-order discontinuity into a smooth crossover on an
$8^3\times
4$ lattice.

A possible interpretation of the smoothing is that it is caused by an explicit
breaking of chiral symmry in the the Wilson quark action, whose effect is
larger for stronger coupling.  If one increases the temporal lattice size
$N_t$,
the cusp and the line of thermal transition will both move toward weaker
coupling.  Therefore the first-order transition may become extended up to and
beyond the tip of the cusp for sufficiently large $N_t$.

Another possibility is that the first-order transition we find is a lattice
artifact, sharing its origin with the sharpening of the
$N_f=2$ transition observed at $\beta\approx 5.0$\cite{milc}.  This possibility
does not contradict the report of a first-order transition for
$N_f=3$\cite{qcdpax}.  In fact
the interval $4.0\leq\beta\leq 4.7$ where first-order signals were observed is
away from the cusp expected at $\beta\approx 3.0$. Also we think that the
divergence of runs at $\beta\approx 3.0$ from a hot and a mixed starting
configurations reported in ref.~\cite{qcdpax} is an indication
that the runs were made inside the parity-broken phase rather than a signal of
metastability.

For either of the two possibilities above, if the chiral phase transition
is of first order for $N_f\geq 3$ in the continuum, it will emerge
only for larger $N_t$ for which the cusp and the thermal line move into the
scaling region toward weak coupling.

\section{Approach toward the continuum limit}

\begin{figure}
\centerline{\epsfxsize=74mm \epsfbox{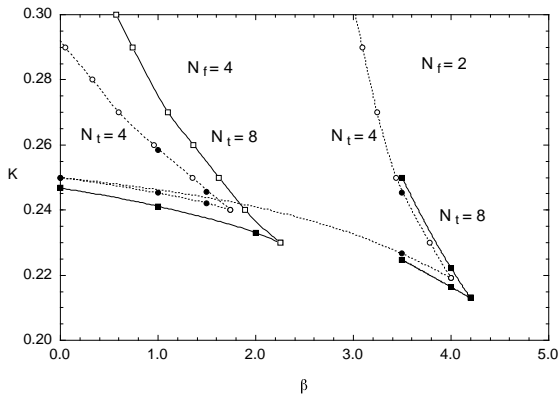}}
\vspace*{-10mm}
\caption{Location of the critical line for the temporal sizes
$N_t=4$ (dotted lines) and 8 (solid lines) for $N_f=2$ and 4.}
\label{fig:fig5}
\vspace*{-8mm}
\end{figure}

Discussions of the previous section naturally raise the question how the
cusp of the critical line moves when the temporal lattice size is increased.
To examine this problem, we analyze the critical line for an $8^3\times
8$ lattice for  $N_f=2$ and 4 systems. We do not repeat a description
of simulation details since they are  similar to those for an $8^3\times 4$
lattice.

In fig.~\ref{fig:fig5} we plot the critical line estimated from the pion
mass.  We observe that the cusp is clearly shifted toward weak coupling.
However, the magnitude of shift is quite small both for $N_f=2$ and 4 so that
the tip of cusp is still in the region of strong coupling for $N_t=8$.
Let us note that previous results by the QCDPAX
Collaboration for $N_f=2$ also indicate that the tip of the cusp is located at
$\beta=4.0-4.2$ for $N_t=6$ and at $\beta=4.5-5.0$ even for
$N_t=18$\cite{qcdpax}.   A recent study on a symmetric lattice employing
parity-flavor breaking external fields also shows that the cusp is located
below
$\beta=5.0$ up to $N_t=10$\cite{bitar}. These results imply that a substantial
increase in the temporal lattice size is needed before the cusp and the thermal
line move into the scaling region ({\it e.g.,}
$\beta\geq 5.5$ for $N_f=2$).

We should emphasize that this result has a significant impact also for
spectrum calculations at zero temperature.  Since the location of the cusp
is controlled by the smaller of the temporal and spatial lattice sizes,
unless the spatial size is taken
sufficiently large, critical hopping parameter will be absent, leading to
systematic errors in measured hadron masses.

Overall our study indicates that within  the plaquette gauge action and
the Wilson quark action large lattice sizes are needed for an exploration
of continuum properties of
the chiral phase transition.  The associated computational difficulties
point toward application of improvement ideas\cite{qcdpaximp,milcimp} to
achieve
further progress in finite-temperature studies with the Wilson-type quark
actions.
\vspace*{1mm}

We thank Y. Iwasaki, K. Kanaya and T. Yoshi\'e for
useful discussions.
This work is supported in part by the Grants-in-Aid of
the Ministry of Education (Nos. 04NP0801, 08640349, 08640350).

\end{document}